\documentclass[twocolumn,aps,prl,superscriptaddress,showpacs]{revtex4}
\usepackage{hyperref}

\usepackage{graphicx}

\usepackage{amsmath}
\usepackage{epsfig}

\begin{document}
 \title{Unconditional quantum cloning of coherent states with linear optics}

\author{Ulrik L. Andersen, Vincent Josse and Gerd Leuchs}
\email{andersen@kerr.physik.uni-erlangen.de}
\affiliation{Institut f\"{u}r Optik, Information und Photonik,  Max-Planck Forschungsgruppe, Universit\"{a}t Erlangen-N\"{u}rnberg, G\"{u}nther-Scharowsky str. 1, 91058, Erlangen, Germany}
\date{\today}

\begin{abstract}
A scheme for optimal Gaussian cloning of optical coherent states is proposed and experimentally demonstrated. Its optical realization is based entirely on simple linear optical elements and homodyne detection. The optimality of the presented scheme is only limited by detection inefficiencies. Experimentally we achieved a cloning fidelity of about 65\%, which almost touches the optimal value of 2/3.
\end{abstract}


\maketitle 

According to the basic laws of quantum mechanics, an unknown nonorthogonal quantum state cannot be copied exactly \cite{wooters82.nat,dieks82.pla}. In other words, it would be an impossible task to devise a process that produces perfect clones of an arbitrary quantum state. However, a physical realization of a quantum cloning machine with less restrictive requirements to the quality of the clones is possible. Such a quantum cloning machine was first considered in a seminal paper by Buzek and Hillery \cite{buzek96.pra} where they went beyond the no-cloning theorem by considering the possibility of producing approximate clones for qubits. These considerations were later extended to the finite-dimensional regime \cite{buzek98.prl} and finally to the continuous variable (CV) regime \cite{cerf00.prl}.
This extension is stimulated by the relative ease in preparing and manipulating quantum states in the CV regime as well as the unconditionalness: Every prepared state is used in the protocols. Governed by these motivations many quantum protocols have been experimentally realized in this regime \cite{vanloock04.xxx}.

Studies on quantum cloning were initially motivated by the apparent implications on quantum information processing but also because they opened an avenue for a clearer understanding of the fundamental concepts of quantum mechanics and measurement theory. Recently, however it has been shown that quantum cloning might improve the performance of some quantum computational task \cite{galvao00.pra} and it is believed to be the optimal eaveasdropping attack for a certain class of quantum key distribution protocols employing coherent states and CV detection \cite{grosshans04.prl}. Furthermore, quantum cloning also provides a means of partial covariant distribution of quantum information between two (or more) parties in a quantum network \cite{braunstein01.pra}.

To date, convincing experimental realizations of quantum cloning have been restricted to the two-dimensional qubit regime where the polarization state of single photons has been conditionally cloned~\cite{linares02.sci,fasel02.prl}. In parallel there have been some theoretical proposals for the experimental implementations of quantum cloning of CV Gaussian states of light~\cite{dariano01.prl,fiurasek01.prl,braunstein01.prl}. These protocols have been shown to be optimal for the Gaussian N$\rightarrow$M cloner where M identical copies are produced from N originals~\cite{cerf00.pra} - a special case being the Gaussian 1$\rightarrow$2 cloner~\cite{nongaussian}. However all these proposals are based on at least one parametric amplifier rendering the practical realization quite difficult. In this Letter we propose a new simple scheme of an optimal Gaussian cloning machine, which does not rely on any non linear interaction but is only based on simple unitary beam splitter transformations and homodyne detection. Furthermore we implemented this idea experimentally for the 1$\rightarrow$2 cloner.      

\begin{figure}[h] \centering \includegraphics[width=7cm]{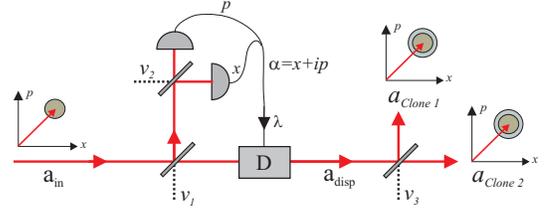} \caption{\it Schematic drawing showing the principle of the continuous variable cloning scheme using only linear optics and homodyne detection. $v_1$, $v_2$ and $v_3$ are vacuum inputs, D is a displacement governed by the measurement $\alpha$ scaled with the gain $\lambda$.}
\label{fig1}  \end{figure}

In this Letter we will consider coherent states of light, for which two canonical conjugate quadratures characterizing the state - e.g. the amplitude, $\hat{x}$, and phase, $\hat{p}$ - have Gaussian statistics. The unknown coherent state to be cloned is then uniquely described by $|\alpha_{in}\rangle=|\frac{1}{2}(x_{in}+ip_{in})\rangle$ where $x_{in}$ and $p_{in}$ are the expectation values of $\hat{x}_{in}$ and $\hat{p}_{in}$. The outputs of the cloning machine are Gaussian mixed states with the expectation values $x_{clone}$ and $p_{clone}$ and characterized by the density operator $\rho_{clone}$. The efficiency of the cloning machine is typically quantified by the fidelity, which gauges the similarity between an input state and an output state. It is defined by $F=\langle \alpha_{in} |\rho_{clone} |\alpha_{in}\rangle$~\cite{furusawa98.sci} and for the particular case of unity cloning gains (corresponding to $x_{clone}=x_{in}$ and $p_{clone}=p_{in}$) it reads
\begin{equation}
F= \frac{2}{\sqrt{(1+\Delta^2x_{clone})(1+\Delta^2p_{clone})}}
\label{fid}
\end{equation}    
where $\Delta^2x_{clone}$ and $\Delta^2p_{clone}$ denote the variances. 

A straight forward way to produce approximate clones uses a measure-and-prepare strategy~\cite{braunstein00.mod,grosshans01.pra}. In such a "classical" scenario, the best approach to cloning an arbitrary coherent state is to measure simultaneously both quadratures $\hat{x}_{in}$ and $\hat{p}_{in}$~\cite{arthur,hammer04.xxx} and subsequently, based on the outcomes of this measurement, clones of the input state are constructed. However, using this procedure two additional units of quantum noise are added to the clones partly due to the attempt to measure two non-commuting variables simultaneously and partly due to the construction of the clones. Although this method enables the production of an infinite number of clones (1$\rightarrow \infty$ cloner), the optimal fidelity is limited to 1/2~\cite{furusawa98.sci,braunstein00.mod,grosshans01.pra,hammer04.xxx}.  

In contrast our quantum approach to cloning uses intrinsic correlations, and runs as follows (see Fig. 1). At the input side of the cloning machine the unknown quantum state is divided by a 50/50 beam splitter. At one output we perform an optimal estimation of the coherent state: the state is split at another 50/50 beam splitter and the amplitude and the phase quadratures are measured simultaneously using ideal homodyne detection~\cite{arthur,hammer04.xxx}.
According to the measurement outcomes the other half of the input state is displaced with a scaling factor, $\lambda$~\cite{lam97.prl}. Using the Heisenberg representation, the displaced field can be expressed as
\begin{eqnarray} 
\hat{a}_{disp}=(\frac{1}{\sqrt{2}}+\frac{\lambda}{2})\hat{a}_{in}+(\frac{1}{\sqrt{2}}-\frac{\lambda}{2})\hat{v}_1-\frac{\lambda}{\sqrt{2}}\hat{v}_2^{\dagger}
\end{eqnarray}
where $\hat{v}_1$ and $\hat{v}_2$ refer to the annihilation operators associated with the uncorrelated vacuum modes entering the two beam splitters (see Fig. \ref{fig1}), and $\hat{a}_{in}$ and $\hat{a}_{disp}$ are the annihilation operators for the input and displaced states.  
In a final step the displaced state is separated in two clones by a 50/50 beamsplitter: 
\begin{eqnarray} 
\hat{a}_{clone1}&=&\hat{a}_{in}+\frac{1}{\sqrt{2}}(\hat{v}_3-\hat{v}_2^{\dagger})\nonumber\\
\hat{a}_{clone2}&=&\hat{a}_{in}-\frac{1}{\sqrt{2}}(\hat{v}_3+\hat{v}_2^{\dagger})
\label{clone_transfer}
\end{eqnarray}
where $\hat{v}_3$ is uncorrelated vaccum noise entering the last beam splitter and $\lambda$ has been taken to be $\sqrt{2}$ to assure unity gain.
The transformations in Eq. (\ref{clone_transfer}) are known to describe an optimal Gaussian cloning machine \cite{fiurasek01.prl,braunstein01.prl}. In particular we see that it is invariant with respect to rotation and displacement in phase space as required by a phase independent or covariant cloner.   
Normalizing the variance of the vacuum state to unity, the variances of the clones for the amplitude and phase quadratures are $\Delta^2x_{clone}=\Delta^2x_{in}+1$ and $\Delta^2p_{clone}=\Delta^2p_{in}+1$, respectively. Note that using the quantum approach only one unit of quantum noise is added in contrast to the classical approach where two units are added. 
Using Eq.(\ref{fid}) the fidelity is found to be 2/3 which corresponds to the optimal fidelity for a Gaussian cloning machine~\cite{cerf00.pra}.

\begin{figure}[h] \centering \includegraphics[width=8cm]{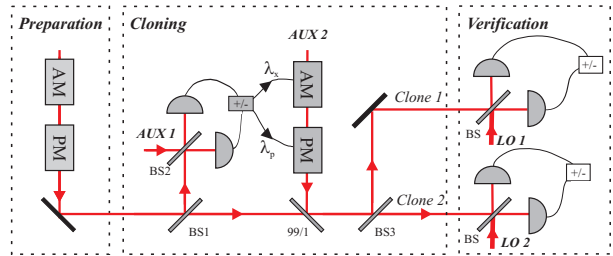} \caption{\it Schematic of the CV cloning setup divided into three boxes defining the preparation stage (where an arbitrary input coherent state can be generated), the cloning stage (where two clones are produced) and the verification stage (where the quality of the cloning process is quantified). BS: Beam splitter, $\lambda$: Electronic gain, LO: Local oscillator, AM: Amplitude modulator, PM: Phase modulator and AUX: Auxiliary beam.}
\label{fig2}  \end{figure}

We now proceed by discussing the experimental demonstration of the proposed scheme. First we present the experimental setup shown in Fig.~2. The laser source for our experiment was a monolithic Nd:YAG(yttrium aluminum garnat) nonplanar ring laser at 1064nm delivering 500mW of power in a single transverse mode. A small part of the power was used to create an input signal to the cloning machine whereas the rest served as local oscillator beams and auxiliary beams. The setup comprises three parts; a preparation stage, a cloning stage and finally a verification stage.

{\it Preparation:} 
In our experiment, we define the quantum state to be frequency sidebands at $\pm$14.3~MHz (with a bandwidth of 100kHz) of a bright electro-magnetic field (similar to previous realisations of CV quantum protocols~\cite{furusawa98.sci}). At this frequency the laser was found to be shot noise limited, ensuring a pure coherent input state. An arbitrary input state is then easily generated by independently controlling the modulations of the amplitude quadrature ($x_{in}$) and the phase quadrature ($p_{in}$), using two electro-optical modulators.

{\it Cloning:} The prepared state is then directed to the cloning machine where it is divided into two halves by the first beam splitter (BS1). One of the halves was combined with an auxiliary beam (AUX1) at the second beam splitter (BS2) with a $\pi/2$ relative phase shift and balanced intensities. The two beam splitter outputs are detecteted by high quantum efficiency photodiodes so that the sum (difference) of the photo currents provide a measure of the amplitude (phase) quadrature of the two beam splitter outputs. This corresponds to an optimal coherent state measurement and therefore a simpler alternative to the one shown in Fig.\ref{fig1}~\cite{leuchs99.mod}. The added and subtracted photocurrents are scaled appropriately with electronic gains $\lambda_x$ and $\lambda_p$ to ensure unity cloning gains, and used to modulate the amplitude and phase of an auxiliary beam (AUX2) via two independent modulators. This beam is then combined at a 99/1 beam splitter with the other half of the signal beam, hereby displacing this part according to measurement outcomes \cite{furusawa98.sci}. 
In a final step, the clones are generated at the output of the third beam splitter (BS3).

{\it Verification:} To characterize the performance of the cloning machine, the spectral noise properties of the two clones are measured by two homodyne detectors with strong local oscillator beams (LO1 and LO2). Since the statistics of the involved light fields are Gaussian we need only measure two conjugate quadratures to fully characterise the states. Therefore the homodyne detectors were set to measure stably - employing electronic servo feedback loops - either the amplitude or the phase quadrature. We note that the input state is also measured by the same homodyne detectors, to ensure a consistent comparison between the input state and the clones.

\begin{figure}[h] \centering \includegraphics[width=8cm]{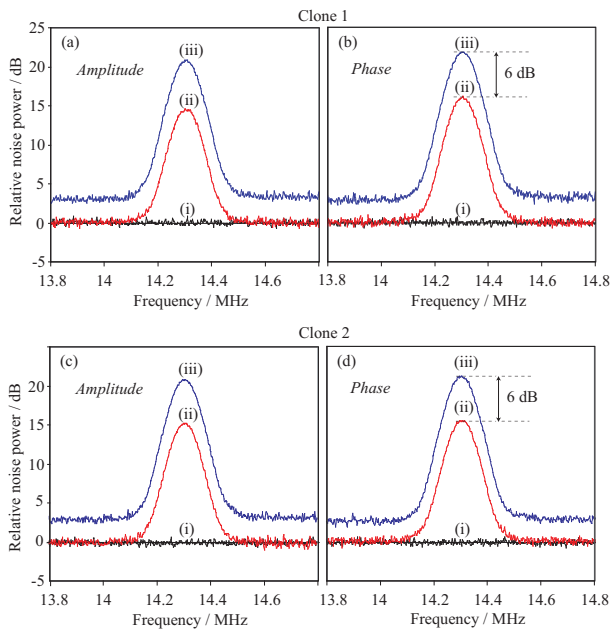} \caption{\it Spectral densities for the input state (traces (ii)) and the output clones (traces (iii)), 1 and 2, of the Gaussian cloning machine, measured by the two homodyne detection systems in a 2 MHz span mode and a center frequency of 14.3MHz corresponding to the location of the sidebands. The traces are normalised to the quantum noise limit (i). Resolution bandwidth is 100~kHz and video band width is 30~Hz. (a) and (b) show the spectra for the amplitude and the phase quadratures of clone 1 and similarily (c) and (d) the spectra for clone 2. Since the input signal was measured with the same homodyne detectors as the output clones, the measurement of the input signal is degraded by the two 50/50 beam splitters BS1 and BS3 and unity cloning gain is ensured by a 6~dB difference between the input signal and the clones.}
\label{fig3}  \end{figure}

An example of a cloning run is reported in Fig.~3. The spectral densities of the amplitude and phase quadratures are here shown over a 1 MHz frequency span for the input state (ii) and the two clones (iii). 
From these traces the coherent amplitude of the various fields, $x_{in,out}$ and $p_{in,out}$, are measured by the heights of the peaks at 14.3~MHz relative to the quantum noise level (i). Using these signal powers we estimate an average photon number of 62 per unit bandwidth per unit time~\cite{photonno}. As evident from the figure, the electronic gains of the feed forward loops are adjusted such that the cloning gains are close to unity (which corresponds to a 6~dB difference between the measured input signal and the output signals due to the degradation of the input signal by BS1 and BS3). In order to simplify the following analysis of the measurement data we will assume unity gains and will later consider the consequences of small deviations from unity which is the case for real cloning machines. 
From Fig.~3 it is also evident that additional noise has been added to the clones relative to the input state which is a result of the cloning action.
In order to quantify accurately the performance of the cloning machine, we estimated precisely this amount of added noise at 14.3 MHz (in a 100kHz bandwidth). To do so, we switched off the modulations of the input beam, and recorded the noise in a zero span measurement over 2 seconds. These results are displayed in Fig.~4 where the added noise in amplitude and phase are reported for both clones. To avoid an erroneously underestimation of the noise power, the traces are corrected to account for the detection efficiencies of the two homodyne stations (which amount to 78.5\% and 77.5\%). From these data, the fidelities of the two copies can be easily determined using Eq.~\ref{fid} and are found to be 64.3$\pm 1\%$ (clone 1) and 65.2$\pm 1\%$ (clone 2), assuming unity cloning gains. These values clearly demonstrate successfull operation of our cloning machine since they significantly surpass the maximum classical fidelity of 50\% and approach the optimal value of 2/3$\approx$66.7\%.

The performance of our system is limited solely by imperfections of the in-line feedforward loop, which include non-unity quantum efficiency of the diodes, electronic noise of the detector circuit and non-perfect interference contrast at the beam splitter BS2 in Fig.~2. The electronic noise was completely overcome by using newly designed ultra low noise detectors (with electronic noise 25dB below the shot noise level)~\cite{bruno}, and the detection efficiencies were maximized by optimizing the mode matching at the beam splitter (99\%) and by using high quantum efficiency photo diodes (95\%). 
Based on these efficiencies we calculate an expected fidelity of 65\% which is in nice agreement with our experimental results. Note that despite of the imperfect detection system, the fidelity is still close to the optimum of 2/3, proving the robustness of the cloning scheme.

\begin{figure}[h] \centering \includegraphics[width=8cm]{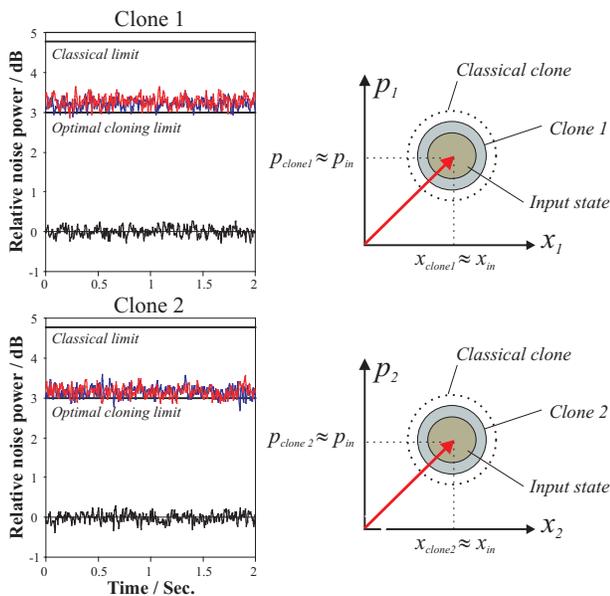} \caption{\it Spectral noise densities of the clones relative to the quantum noise level (black trace) recorded by the homodyne detectors both for the amplitude quadrature (red trace) and for the phase quadrature (blue trace). The added noise contributions are 3.28$\pm 0.13$~dB (3.16$\pm 0.13$~dB) and 3.20$\pm 0.11$~dB (3.15$\pm 0.13$~dB) in the amplitude quadrature and phase quadrature of clone 1(2). The optimal cloning limit as well as the classical limit are shown by solid lines.  
The measurement frequency is 14.3~MHz, the sweep time 2 seconds, the resolution bandwidth 100~kHz and the video bandwidth 300~Hz. On the right hand side we plot the associated noise contours of the Wigner functions corresponding to the input state (green contour), the experimentally achieved clones (light blue contour) and the classical clones (dashed line).}
\label{fig4}  \end{figure}

In the discussion above we assumed unity gains. However, experimental imperfections lead to a small deviation from unity, and the gains were accurately determined to be $g_{x1}=0.96\pm0.01$ and $g_{p1}=1.00\pm0.01$ for clone 1 and $g_{x2}=1.03\pm0.01$ and $g_{p2}=1.03\pm0.01$ for clone 2 for the amplitude and phase quadratures, respectively. 
As a result of the deviations from unity gain, the fidelity depends on the photon number of the input coherent state and the figure of merit for cloning is an average of the "single-shot" fidelities~\cite{cochrane04.pra}. E.g. considering a Gaussian distributed set of input coherent states with a spread in photon number of 50 (which is a huge number in quantum information science) the average fidelities equal 62.7\% and 63.3\%, which still significantly exceed the classical cloning boundary of 50.2\% for the same span of input states. 
Despite the fact that the gains are not exactly unity, the obtained fidelities are far above the classical limits and approach the optimal limits for a large set of input states, demonstrating the suitability of this cloning machine for realistic experimental quantum information tasks.

In conclusion, we have proposed a simple Gaussian cloning protocol based on linear optics and homodyne detection which is optimal for coherent state inputs, and we have experimentally demonstrated the idea and obtained near optimal unconditional quantum cloning of coherent states.
Finally, let us stress that it is straightforward to extend the presented scheme for 1$\rightarrow$2 cloning (using linear optics) to a large variety of different copying functions such as optimal N$\rightarrow$M Gaussian cloning function which takes N originals and produces M clones~\cite{cerf00.pra,fiurasek01.prl,braunstein01.prl} and an asymmetric cloning function which produces output clones of different quality~\cite{fiurasek01.prl}, a procedure which is crucial in controlled partial information transfer between different parties in a network 

We thank T.C. Ralph, R. Filip, N. Treps, W. Bowen, R. Schnabel and J. Sherson for stimulating discussions and B. Menegozzi for the construction of the photodetectors. This work has been supported by DFG (the Schwerpunkt programm 1078), the network of competence QIP (A8), and EU projects COVAQIAL (project no. FP6-511004) and SECOQC. ULA acknowledges an Alexander von Humboldt fellowship. 

\bibliography{scibib}

\begin{thebibliography}{}

\bibitem{wooters82.nat} W.K. Wooters and W.H. Zurek, Nature {\bf 299}, 802 (1982). 
\bibitem{dieks82.pla} D. Dieks, Phys. Lett. A {\bf 92}, 271 (1982).
\bibitem{buzek96.pra} V. Buzek and M. Hillery, Phys. Rev. A {\bf 54}, 1844 (1996).
\bibitem{buzek98.prl} V. Buzek and M. Hillery, Phys. Rev. Lett. {\bf 81}, 5003 (1998).
\bibitem{cerf00.prl} N. Cerf et al.,  Phys. Rev. Lett. {\bf 85}, 1754 (2000).
\bibitem{vanloock04.xxx} S. Braunstein and P. van Loock. To appear in Rev. Mod. Phys. http://lanl.arxiv.org/abs/quant-ph/0410100
 \bibitem{galvao00.pra} E.F. Galvao and L. Hardy, Phys. Rev. A {\bf 62} 022301 (2000).
\bibitem{grosshans04.prl} F. Grosshans and N. Cerf, Phys. Rev. Lett. {\bf 92}, 047905 (2004).
\bibitem{braunstein01.pra} S. Braunstein et al., Phys. Rev. A {\bf 63}, 052313 (2001).
\bibitem{linares02.sci} A. Lamas-Linares et al., Science {\bf 296}, 712 (2002).
\bibitem{fasel02.prl} S. Fasel et al., Phys. Rev. Lett. {\bf 89}, 107901 (2002).
\bibitem{dariano01.prl} G.M. D'Ariano et al., Phys. Rev. Lett. {\bf 86}, 914 (2001).
\bibitem{braunstein01.prl} S.L. Braunstein et al., Phys. Rev. Lett. {\bf 86}, 4938 (2001).
\bibitem{fiurasek01.prl} J. Fiurasek, Phys. Rev. Lett. {\bf 86}, 4942 (2001).
\bibitem{cerf00.pra} N. Cerf and S. Iblisdir, Phys. Rev. A {\bf 62}, 040301(R) (2000).
\bibitem{nongaussian} Recently it has been shown in ref. [N.J. Cerf, O. Kr\"u ger et al. quant-ph/0410058] that the optimal coherent state cloner is non-Gaussian. The optimal fidelity is approximately 0.6825 which is slightly higher than the
optimal fidelity of 2/3 in the Gaussian setting.
\bibitem{furusawa98.sci} A. Furusawa et al., Science {\bf 282}, 706 (1998).
\bibitem{braunstein00.mod} S.L. Braunstein et al. J. Mod. Opt. {\bf 47} 267 (2000).
\bibitem{grosshans01.pra} F. Grosshans and Ph. Grangier, Phys. Rev. A {\bf 64}, 010301(R) (2001).
\bibitem{arthur} E. Arthur and J.L. Kelly, Bell Syst. Technol. {\bf 44}, 725 (1965). 
\bibitem{hammer04.xxx} K. Hammerer et al. http://lanl.arxiv.org/abs/quant-ph/0409109
\bibitem{lam97.prl} P.K. Lam et al. Phys. Rev. Lett. {\bf 79}, 1471 (1997).
\bibitem{leuchs99.mod} G. Leuchs et al. Jour. Mod. Opt. {\bf 46}, 1927 (1999).
\bibitem{bruno} B. Menegozzi et al. manuscript in preparation
\bibitem{photonno} The average number of photons per unit bandwidth per unit time in a sideband of the optical field is calculated using the expression $\langle n\rangle =(\Delta ^2x+\Delta ^2p-2)/4$. 
\bibitem{cochrane04.pra} P.T. Cochrane et al. Phys. Rev. A {\bf 69}, 042313 (2004).


\end{thebibliography}

\bibliographystyle{Science}

\end{document}